\documentclass[final,5p,times]{elsarticle}

\usepackage{amsmath}
\usepackage{amssymb}
\usepackage{mathtools}
\usepackage{graphicx}
\usepackage{booktabs}
\usepackage{array}
\usepackage{dcolumn}
\usepackage{xcolor}
\usepackage{hyperref}
\usepackage{tikz}
\usepackage{chngcntr}
\usetikzlibrary{calc}

\biboptions{numbers,sort&compress}

\newcolumntype{d}[1]{D{.}{.}{#1}}

\hypersetup{hidelinks}

\begin{document}

\begin{frontmatter}

\title{Whole-Piece Training for Symbolic Music Language Models via Full-Horizon Compressed Recurrence}

\author[inst1]{Yungang Yi\corref{cor1}}
\ead{yungang.yi@aut.ac.nz}
\cortext[cor1]{Corresponding author.}

\author[inst1]{Weihua Li}
\ead{weihua.li@aut.ac.nz}

\author[inst1]{Matthew Kuo}
\ead{matthew.kuo@aut.ac.nz}

\author[inst1]{Catherine Shi}
\ead{catherine.shi@aut.ac.nz}

\author[inst2]{Quan Bai}
\ead{quan.bai@utas.edu.au}

\address[inst1]{Auckland University of Technology, Auckland, New Zealand}
\address[inst2]{University of Tasmania, Tasmania, Australia}

\begin{abstract}

For computational efficiency, modern language models are typically trained on independently sampled fixed-length sequences. Symbolic music language models largely inherit this paradigm, despite musical structure naturally unfolding over complete compositions rather than isolated excerpts. Fragmenting compositions into independent training instances therefore prevents continuous conditioning over the complete work.

We present a practical framework for whole-piece training of symbolic music language models via Full-Horizon Compressed Recurrence (FHCR). FHCR preserves the full temporal horizon of recurrent memory while reducing
the dimensionality of its key-value (KV) representation, making continuous
whole-piece training practical under limited GPU memory.

To directly assess functional long-range dependence, we introduce KV-Reset Context Utilization (KRCU), an evaluation-time diagnostic. On the MAESTRO symbolic piano dataset, KRCU shows that full-horizon models utilize context far beyond the local segment window, whereas reducing the temporal extent of recurrent memory substantially
weakens this measurable long-range dependence. FHCR preserves long-range context utilization while substantially reducing recurrent memory cost.

These findings show that preserving the temporal extent of recurrent history
is important for efficient whole-piece modeling, and that memory cost can
instead be reduced through KV representation compression.

The project demos and generated music samples are available at \url{https://wholemusic.github.io}.

\end{abstract}

\begin{keyword}
Symbolic music generation \sep Whole-piece training \sep Long-context modeling \sep Full-Horizon Compressed Recurrence \sep Context utilization
\end{keyword}

\end{frontmatter}

\section{Introduction}

Modern language models are typically trained on independently sampled fixed-length sequences for computational efficiency. Symbolic music language models largely inherit this training paradigm, despite musical structure naturally unfolding over complete compositions rather than isolated excerpts. Motif recurrence, phrase development, harmonic progression, and large-scale musical form can span thousands of symbolic events or even an entire composition \cite{huang2019musictransformer,yu2022museformer}. Fragmenting compositions into independent training instances therefore removes cross-segment musical dependencies that naturally exist within the original work.

A natural alternative is \textbf{whole-piece training}, in which each complete composition is preserved as a single continuous training instance while computation is streamed over shorter recurrent segments. Segment-level recurrence, introduced by Transformer-XL \cite{dai2019transformerxl}, provides a natural mechanism for realizing this formulation by propagating recurrent states across segment boundaries. In this setting, segmentation becomes a computational device rather than the statistical boundary of a training sample, allowing predictions later in a composition to remain conditioned on information originating much earlier in the same work.

The challenge is that extending recurrence from a limited local history to a complete composition is expensive. Transformer-XL demonstrated segment-level recurrence with training-time memory typically comparable to a local segment \cite{dai2019transformerxl}, rather than histories spanning many recurrent steps. Preserving complete recurrent history requires maintaining cross-segment key-value (KV) states over long temporal horizons and across Transformer layers, causing recurrent-state memory to grow rapidly with composition length. Whole-piece training therefore raises a fundamental efficiency question: how can recurrent memory be reduced without destroying the long-range information that motivates whole-piece training in the first place?

A straightforward solution is to reduce the temporal extent of recurrent memory by truncating historical states or retaining long histories only at selected depths. However, whether such memory reduction actually preserves functional long-range dependence cannot be determined from aggregate perplexity alone. To measure this directly, we introduce \textbf{KV-Reset Context Utilization (KRCU)}, an evaluation-time diagnostic that intervenes on recurrent KV states while keeping model parameters fixed and measures the resulting change in future-token negative log-likelihood. KRCU therefore tests whether information carried from before the reset boundary was functionally contributing to subsequent predictions.

KRCU reveals an important distinction. Full-horizon recurrence exhibits
measurable dependence on context far beyond the local segment window,
whereas reducing the temporal extent of recurrent memory can substantially
weaken this dependence despite competitive aggregate perplexity. This suggests
that temporal memory reduction may remove precisely the information that
whole-piece training is intended to preserve.

Motivated by this observation, we present \textbf{Full-Horizon Compressed
Recurrence (FHCR)}, which preserves the complete temporal horizon while
reducing the dimensionality of the recurrent KV representation. Building on
grouped-query attention (GQA), FHCR extends KV representation reduction from
autoregressive inference to training-time recurrence. Experiments on MAESTRO
show that FHCR preserves measurable long-range context utilization while
substantially reducing recurrent memory cost.

Our contributions are threefold. First, we formulate whole-piece training for symbolic music language models, preserving each complete composition as a continuous training instance through streamed recurrent computation. Second, we introduce KV-Reset Context Utilization (KRCU), an evaluation-time diagnostic for functional long-range dependence, and use it to show that temporal memory reduction can substantially weaken measurable long-range
context utilization despite competitive aggregate perplexity. Third, we introduce Full-Horizon Compressed Recurrence (FHCR), which preserves the complete temporal horizon while reducing recurrent KV-head dimensionality, providing a practical alternative to temporal memory truncation for memory-efficient whole-piece training.

The remainder of this paper is organized as follows.
Section~\ref{sec:related_work} reviews related work.
Section~\ref{sec:method} presents whole-piece training and Full-Horizon Compressed Recurrence.
Section~\ref{sec:exp} describes the experimental setup and KRCU protocol.
Section~\ref{sec:results} reports modeling efficiency and long-range context-utilization results.
Section~\ref{sec:discussion} discusses the implications of preserving temporal coverage while reducing recurrent representation cost.

\section{Related Work}
\label{sec:related_work}

We review related work from five perspectives: (1) training fragmentation and whole-sequence learning, (2) long-context symbolic music modeling, (3) recurrent streaming and temporal memory pruning, (4) diagnosing long-range context utilization, and (5) KV representation compression.

\subsection{Training Fragmentation and Whole-Sequence Learning}
\label{sec:rw_training}

Modern language models are typically trained on fixed-length sequences for computational efficiency, often fragmenting naturally coherent documents into separate training instances. Recent work has begun to question this practice. ERNIE-Doc explicitly addresses context fragmentation through document-level pretraining with recurrent cross-segment context~\cite{ding2021erniedoc}. Zhao et al.\cite{zhao2024sequencecomposition} show that sequence composition strategies significantly influence language-model pretraining, while Ding et al.\cite{ding2024fewertruncations} demonstrate that unnecessary document truncation can degrade language modeling even when complete documents fit within the available context window.

Symbolic music presents a particularly natural setting for preserving complete training instances, since motifs, phrase development, harmonic progression, and sectional form can span an entire composition. Nevertheless, symbolic music language models largely inherit excerpt-based training from language modeling. Our work instead considers whole-piece training, where each complete composition remains a continuous training instance while computation is streamed recurrently.

\subsection{Long-Context Symbolic Music Modeling}
\label{sec:rw_music}

Long-range musical coherence has motivated a variety of architectural approaches. Music Transformer improves relative positional representations for long-range dependency modeling~\cite{huang2019musictransformer}, while Museformer introduces fine--coarse attention to balance local and global musical structure~\cite{yu2022museformer}. WholeSong adopts hierarchical diffusion for complete-song generation~\cite{wang2024wholesong}, MuPT explores long-context pretraining for symbolic music foundation models~\cite{qu2025mupt}, and large-scale autoregressive models such as NotaGen further improve musical modeling through large-scale pretraining~\cite{wang2025notagen}.

These approaches increase the architectural capacity for modeling long musical contexts, but generally rely on bounded training contexts or hierarchical decomposition rather than preserving each composition as a continuous recurrent training instance. Our work is complementary: rather than introducing a new musical sequence operator, we investigate how complete compositions can be continuously modeled during training under practical memory constraints.

\subsection{Recurrent Streaming and Temporal Memory Pruning}
\label{sec:rw_recurrence}

Recurrent Transformers provide a natural mechanism for streaming long sequences while maintaining cross-segment context. Transformer-XL introduced segment-level recurrence by carrying cached representations across segments~\cite{dai2019transformerxl}. Subsequent approaches extend the effective recurrent history through alternative memory representations. Compressive Transformer compresses older memories into lower-resolution representations~\cite{rae2020compressive}; Recurrent Memory Transformer carries a fixed set of learned memory tokens across segments~\cite{bulatov2022rmt}; and ERNIE-Doc extends recurrent document modeling through enhanced cross-segment information flow~\cite{ding2021erniedoc}.

A more direct strategy for reducing memory cost is to prune temporal access by limiting how much historical context is retained. Transformer-XL itself uses a fixed recurrent horizon, while layer-dependent memory spans~\cite{rae2020deepmemory} and architectures that interleave local and full attention~\cite{mimo2026} further reduce where long-range history remains directly accessible. Such approaches reduce memory or computation by restricting the temporal extent or placement of long-range context.

Our experiments examine a complementary question: whether such temporal memory reduction preserves functional long-range dependence rather than merely aggregate predictive performance. This distinction motivates the context-utilization diagnostic described next.

\subsection{Diagnosing Long-Range Context Utilization}
\label{sec:rw_context_utilization}

The availability of a long context window does not necessarily imply that a model functionally uses distant information. Prior work has therefore examined long-context utilization by varying the amount or location of input context available at evaluation time and measuring the resulting changes in prediction behavior~\cite{sun2021longrange}. Related context-ablation analyses similarly assess the contribution of particular portions of the input by removing them and observing changes in model outputs.

Our setting differs because long-range information is carried through recurrent KV states rather than repeatedly supplied as an explicit input prefix. We therefore introduce KV-Reset Context Utilization (KRCU), which intervenes directly on the recurrent state at evaluation time while keeping model parameters fixed, and measures the resulting change in future-token negative log-likelihood. KRCU tests whether information propagated from before the reset boundary remains functionally relevant to subsequent predictions.

KRCU therefore complements perplexity by distinguishing nominal recurrent
capacity from functional dependence on distant history.

\subsection{KV Representation Compression}
\label{sec:rw_kv_compression}

An alternative to pruning temporal history is to reduce how that history is represented. Multi-query attention (MQA) shares a single key/value head across multiple query heads, substantially reducing KV-cache size and memory bandwidth~\cite{shazeer2019fast}. Grouped-query attention (GQA) generalizes this design by sharing key/value heads within groups, providing a compromise between multi-head attention and MQA while maintaining strong model quality~\cite{ainslie2023gqa}. These techniques have been particularly influential for reducing KV-cache and memory-bandwidth costs during autoregressive inference.

A growing body of work further reduces inference-time KV representation costs. MiniCache exploits similarity between neighboring Transformer layers~\cite{liu2024minicache}, SqueezeAttention assigns non-uniform KV-cache budgets according to layer sensitivity~\cite{wang2025squeezeattention}, and LAVa derives adaptive cache allocation from information-loss considerations~\cite{shen2025lava}. Collectively, these studies demonstrate that the representation cost of historical context can be substantially reduced without simply shortening the context window.

FHCR builds on this principle but applies it to a different setting: training-time recurrent memory for whole-piece modeling. Rather than shortening the recurrent horizon, FHCR preserves the complete temporal history while reducing recurrent-state cost along the KV-head dimension. This distinction---reducing how recurrent history is represented rather than how much history is retained---is central to our approach.

\section{Whole-Piece Training via Full-Horizon Compressed Recurrence}
\label{sec:method}

Whole-piece training separates the statistical definition of a training instance from the computational unit used to process it. We formulate symbolic music language model training as whole-piece training, in which each complete musical composition is preserved as a single continuous training instance rather than fragmented into independently optimized excerpts. Computation is streamed sequentially through recurrent segments, allowing compositions to be processed without performing full-sequence attention or backpropagation through the entire work at once while maintaining continuous conditioning across the composition.

The principal challenge is recurrent-memory cost, which grows with sequence
length, model depth, and the number of KV heads. We address this through
\textbf{Full-Horizon Compressed Recurrence (FHCR)}, described below.

\subsection{Streaming Recurrent Realization}
\label{sec:method_recurrence_overview}

Whole-piece training is realized through Transformer-XL-style segment-level recurrence. Each composition is streamed sequentially as a series of recurrent segments, while cached states propagate contextual information across segment boundaries. Unlike conventional fixed-window training, segmentation serves only as a computational strategy: it determines how the composition is processed, rather than defining independent training instances.

Let a composition be partitioned into recurrent segments
$\{x^{(t)}\}_{t=1}^{T}$.
For each segment $t$, queries from the current segment attend to a context consisting of the current segment together with recurrent states propagated from preceding segments. Thus, information can flow continuously across segment boundaries even though computation proceeds one segment at a time.

\begin{figure}[t]
  \centering
  \includegraphics[width=\linewidth]{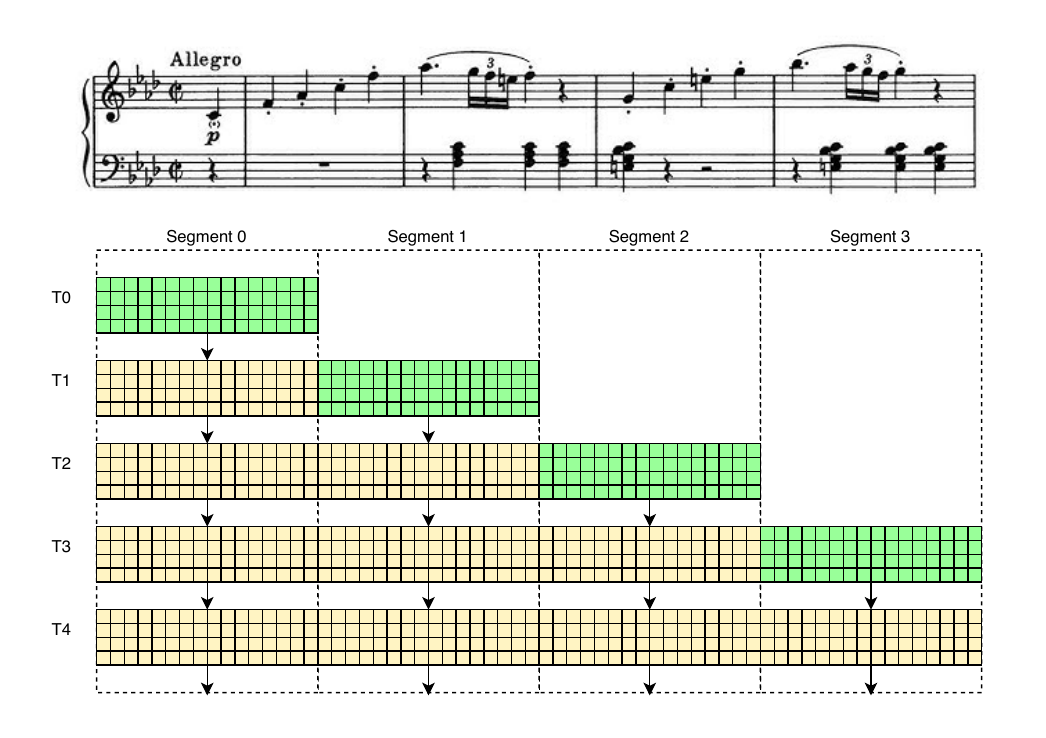}
  \caption{
  \textbf{Whole-piece training through streamed recurrent segments.}
  A complete musical composition remains a single continuous training instance while computation proceeds sequentially over recurrent segments.
  Segmentation is introduced solely for efficient computation rather than defining independent optimization samples.
  }
  \label{fig:full_piece_recurrence}
\end{figure}

Fig.~\ref{fig:full_piece_recurrence} illustrates this streamed realization of whole-piece training.

\subsection{Whole-Piece Sampling}
\label{sec:method_sampling}

During training, each optimization sample corresponds to one complete musical composition. Partitioning a composition into recurrent segments does not alter this statistical training unit; it only determines how computation is streamed.

For each segment $t$, predictions are conditioned on the current segment together with information propagated through recurrent states from preceding segments. Consequently, predictions are not restricted by the boundary of an independently sampled excerpt and can utilize information originating earlier in the same composition.

For data augmentation, we randomize the length of the first segment by sampling its token length from a predefined range, while all subsequent segments use a fixed length. This varies the recurrent segment boundaries observed during training without fragmenting the composition into independent samples.

\subsection{Full-Horizon Recurrent Reference}
\label{sec:ref_full_attn_approx}

We first define a full-horizon recurrent reference, in which every Transformer layer retains recurrent states from the complete available prefix of the composition. Unlike conventional finite-horizon recurrence, no historical states are removed solely because they exceed a fixed temporal window.

For layer $\ell$, let
\[
K^{(\ell)}_{<t}, V^{(\ell)}_{<t}
\]
denote the recurrent key/value states accumulated from preceding segments and
\[
K^{(\ell)}_{t}, V^{(\ell)}_{t}
\]
the states of the current segment. The attention context is formed as
\begin{align}
K^{(\ell)}_{\mathrm{ctx},t}
&=
\left[
K^{(\ell)}_{<t};
K^{(\ell)}_{t}
\right],\\
V^{(\ell)}_{\mathrm{ctx},t}
&=
\left[
V^{(\ell)}_{<t};
V^{(\ell)}_{t}
\right].
\end{align}

The recurrent history therefore grows with the available prefix until the beginning of the composition is reached. This provides full-prefix recurrent conditioning in the forward pass while retaining streamed execution.

As in Transformer-XL-style recurrence, cached states are detached across segment boundaries. Let $\theta$ denote the current parameters and $\theta_{\mathrm{old}}$ the parameters under which a cached state was produced. Then
\[
(K_{\mathrm{cache}},V_{\mathrm{cache}})
=
(K(\theta_{\mathrm{old}}),V(\theta_{\mathrm{old}})),
\]
whereas exact full-sequence Transformer training would recompute historical states under the current parameters. Full-horizon recurrence therefore differs from exact full-sequence attention in gradient propagation and state freshness, while preserving access to the complete available historical prefix during forward computation.

The full-horizon reference provides the unconstrained temporal-memory baseline
for whole-piece conditioning, but its memory cost is substantial because every historical token retains a full KV representation at every layer.

\subsection{Full-Horizon Compressed Recurrence}
\label{sec:method_fhcr}

The central design principle of Full-Horizon Compressed Recurrence (FHCR) is to reduce recurrent-memory cost without shortening its temporal horizon. In whole-piece recurrence, memory cost depends not only on the number of historical positions retained, but also on the number of key/value (KV) heads stored for each position. Temporal pruning reduces the former by discarding older recurrent states. FHCR instead preserves the complete available recurrent history while reducing the number of KV heads.

We realize FHCR using Grouped-Query Attention (GQA)~\cite{ainslie2023gqa}. GQA divides query heads into groups that share key and value heads, interpolating between Multi-Head Attention (MHA), where each query head has a separate KV head, and Multi-Query Attention (MQA), where all query heads share a single KV head~\cite{ainslie2023gqa}. In our recurrent setting, the same grouping applies to the KV states propagated across segments.

Our full-horizon recurrent reference uses 16 query heads and 16 KV heads. FHCR retains the same 16 query heads but uses only 2 KV heads. Consequently, the complete temporal horizon remains available at every recurrent layer, while substantially fewer KV states must be retained for each historical position.

Thus, FHCR reduces recurrent-state cost without removing historical positions:
temporal coverage is unchanged, while the KV representation stored at each
position is reduced. Whether this structural preservation translates into
functional long-range utilization is evaluated using KRCU in
Sec.~\ref{sec:method_krcu}.

\begin{figure}[t]
  \centering
  \includegraphics[width=\linewidth]{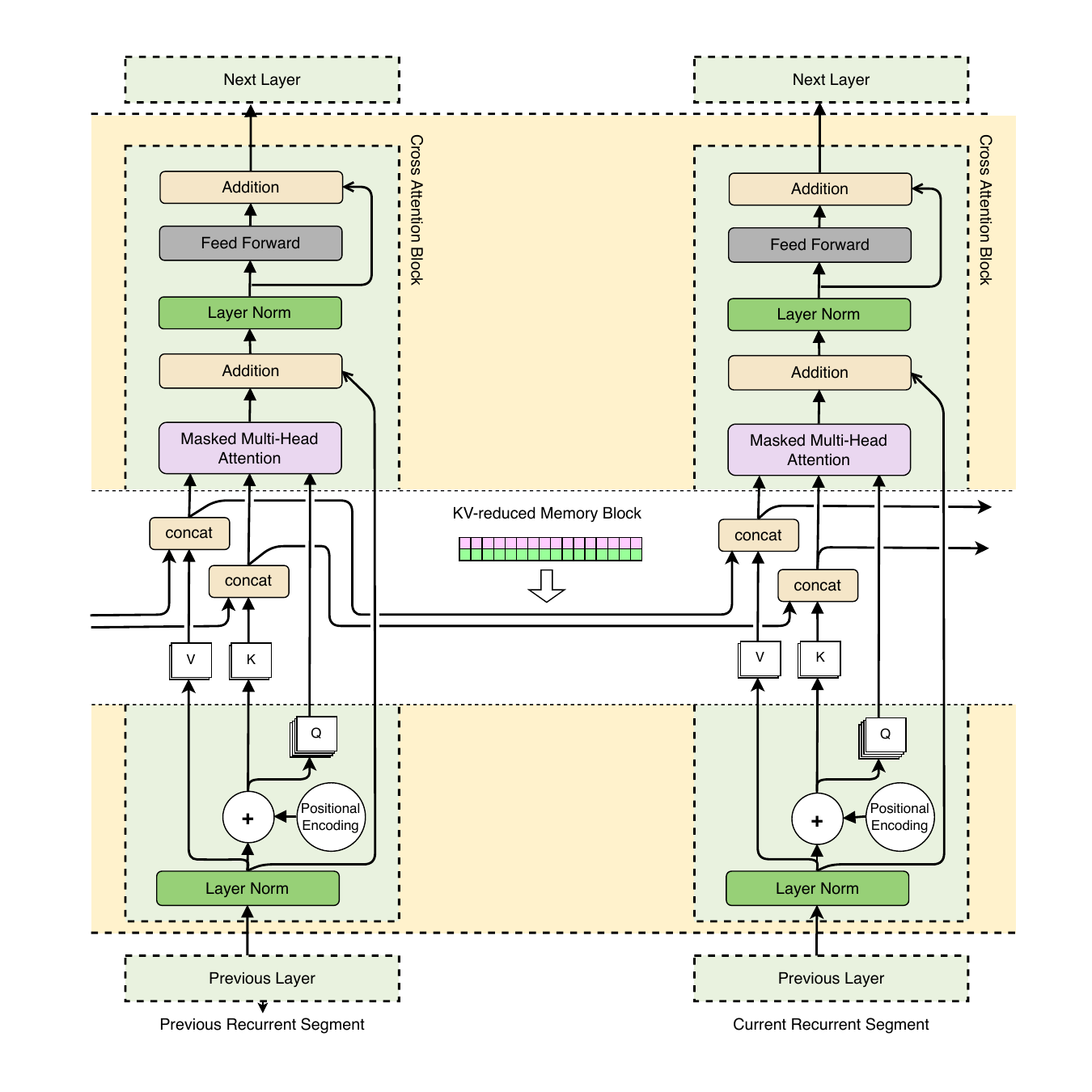}
  \caption{
  \textbf{Full-Horizon Compressed Recurrence (FHCR).}
  Temporal-memory pruning reduces recurrent cost by discarding older historical states, shortening the accessible temporal horizon.
  FHCR instead preserves the complete recurrent horizon while reducing the number of recurrent KV heads through grouped-query attention.
  }
  \label{fig:fhcr}
\end{figure}

Fig.~\ref{fig:fhcr} illustrates this distinction.

\subsection{KV-Reset Context Utilization}
\label{sec:method_krcu}

Architectural access to a long recurrent horizon does not establish that
the model functionally utilizes information over that horizon.
We therefore introduce \textbf{KV-Reset Context Utilization (KRCU)}, an
evaluation-time intervention that measures the predictive effect of
removing previously accumulated recurrent information.

At a controlled boundary $r$, recurrent KV states are reset while model
parameters remain unchanged. Let $d$ denote the distance from the reset
boundary. We define

\begin{equation}
\Delta\mathrm{NLL}(d)
=
\mathrm{NLL}_{\mathrm{reset}}(r+d)
-
\mathrm{NLL}_{\mathrm{full}}(r+d).
\end{equation}

A positive $\Delta\mathrm{NLL}(d)$ indicates that information accumulated
before the reset boundary remains functionally relevant $d$ tokens after
the intervention.

To summarize the magnitude and persistence of the reset effect, we define
KRCU-AUC as the normalized area under the post-reset KRCU curve.
Let $N$ denote the number of equally spaced post-reset evaluation windows
and let $\Delta\mathrm{NLL}_i$ denote the mean NLL increase within the
$i$th window. We define

\begin{equation}
\mathrm{KRCU\text{-}AUC}
=
\frac{1}{N}
\sum_{i=1}^{N}
\Delta\mathrm{NLL}_i.
\label{eq:krcu_auc}
\end{equation}

Because all evaluation windows are equally spaced over the same post-reset
range, Eq.~\eqref{eq:krcu_auc} is the mean post-reset $\Delta$NLL and
equivalently a normalized discrete AUC over that range. Higher KRCU-AUC
indicates a larger and more persistent predictive effect of removing pre-reset
recurrent information.

\section{Experimental Setup}
\label{sec:exp}

Our experiments address three questions motivated by whole-piece training.
First, does full-horizon recurrent training produce models that functionally utilize musical context far beyond the local segment window?
Second, can recurrent memory be reduced by shortening its temporal extent without losing this long-range dependence?
Third, can Full-Horizon Compressed Recurrence (FHCR) instead reduce recurrent-memory cost through fewer KV heads while preserving both the full temporal horizon and functional long-range context utilization?

We evaluate these questions using validation perplexity, computational-efficiency
measurements, and the proposed \textbf{KV-Reset Context Utilization (KRCU)}
analysis (Sec.~\ref{sec:method_krcu}).

\subsection{Dataset and Tokenization}
\label{sec:exp_data}

We conduct experiments on the MAESTRO dataset, a large-scale collection of
virtuosic piano performances with high-quality MIDI recordings~\cite{hawthorne2019maestro}.
We retain compositions whose tokenized lengths fall between 1024 and 32768 tokens.
The lower bound ensures that every retained composition spans at least one recurrent segment,
while the upper bound defines the maximum composition length considered in our experiments.
After applying this length filter, 1,045 of the original 1,276 MAESTRO performances
are retained (81.9\%).
Each retained composition is processed in its entirety rather than truncated into
independent training excerpts.

The resulting train/validation/test splits contain 933, 101, and 11 compositions,
respectively.
We train autoregressive models over symbolic event sequences derived from MIDI.
Each MIDI performance is encoded using note-on, note-off, time-shift, velocity,
and special events, yielding a vocabulary of size $|\mathcal{V}|=393$.

\subsection{Model and Recurrent Configuration}
\label{sec:exp_model}

All models use an 18-layer Transformer backbone with hidden size $d=1024$, 16 query heads, head dimension $d_{\mathrm{head}}=64$, and feed-forward dimension $d_{\mathrm{ff}}=4096$.
Each composition is streamed left-to-right in recurrent segments of length $s=1024$ as described in Sec.~\ref{sec:method_recurrence_overview}.

The maximum recurrent horizon is 31744 tokens, allowing predictions in the final segment of the longest retained composition to receive recurrent information originating from its complete preceding history.

\paragraph{Full-horizon MHA reference.}
Our primary reference retains the complete available recurrent history at every Transformer layer and uses standard multi-head attention (MHA) with 16 query heads and 16 KV heads.
No recurrent states are removed because of their temporal distance from the current segment.
This configuration establishes the long-range context utilization achievable under uninterrupted whole-piece recurrence.

\paragraph{FHCR.}
Our primary memory-efficient model retains the same complete recurrent horizon at every layer but replaces MHA with grouped-query attention (GQA)~\cite{ainslie2023gqa}.
The model retains 16 query heads while using 2 KV heads.
FHCR therefore differs from the full-horizon MHA reference in KV-head dimensionality rather than temporal coverage, isolating whether recurrent-memory cost can be reduced without shortening the available history.

\subsection{Temporal-Memory Pruning Baselines}
\label{sec:exp_temporal_pruning}

\begin{figure}[t]
  \centering
  \includegraphics[width=\linewidth]{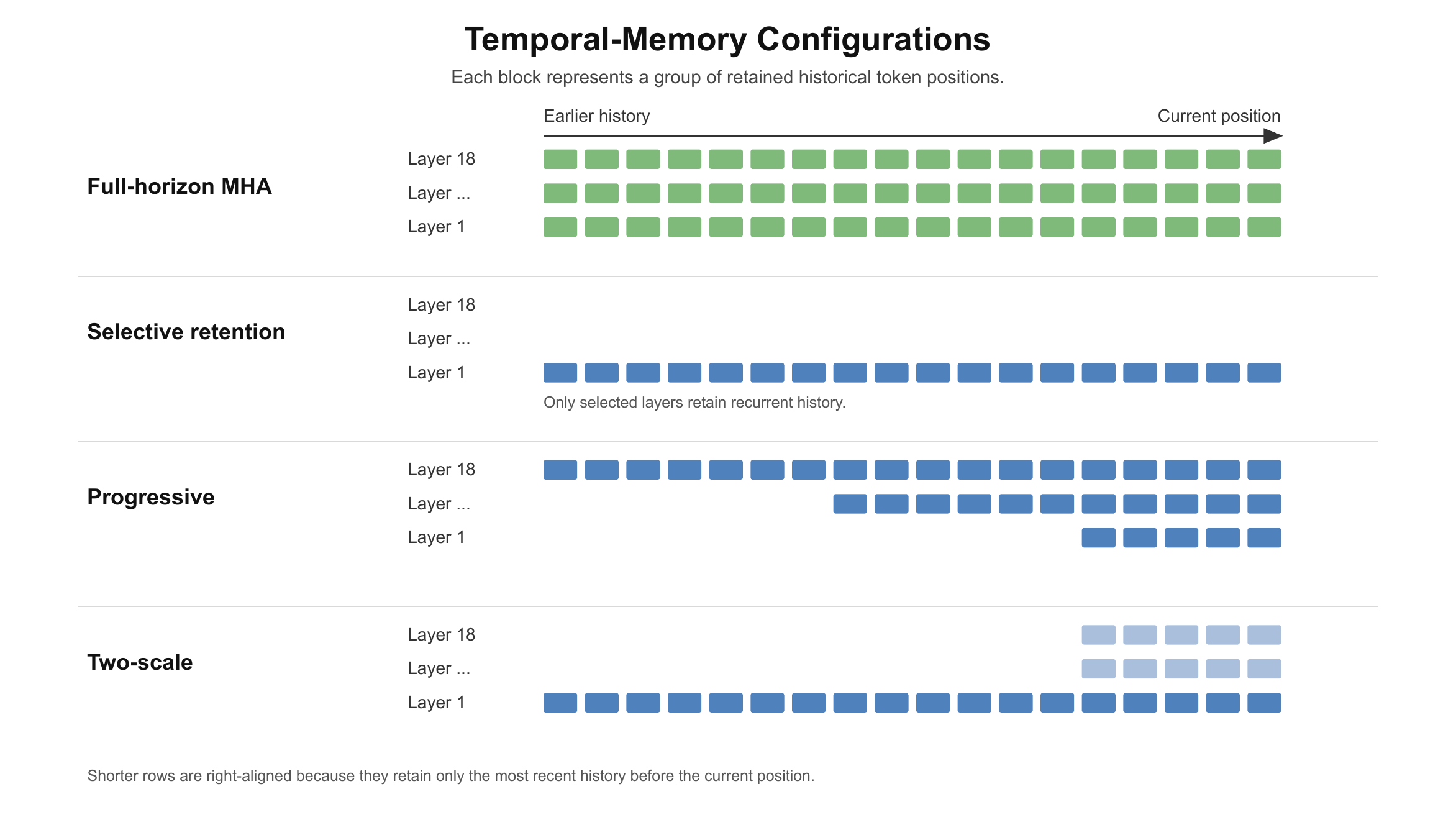}
  \caption{
  \textbf{Temporal-memory configurations used in the experiments.}
  Bar length indicates the recurrent history retained at each Transformer
  depth, with all horizons aligned to the current position.
  The full-horizon reference retains the complete available history at every
  layer, whereas selective retention, progressive allocation, and two-scale
  allocation reduce the temporal extent or depth-wise availability of
  recurrent memory.
  }
  \label{fig:temporal_memory_configurations}
\end{figure}

To distinguish KV-head reduction from temporal-memory reduction, we compare FHCR against several recurrent-memory pruning strategies.
These baselines reduce recurrent cost by shortening or removing recurrent history at selected Transformer layers rather than reducing the number of KV heads.
All use the same Transformer backbone and training procedure as the full-horizon reference. Figure~\ref{fig:temporal_memory_configurations} illustrates the depth-wise recurrent horizons of the principal temporal-memory configurations.

\paragraph{Selective layer retention.}
Only three Transformer layers retain the maximum recurrent horizon, while the remaining layers have no cross-segment recurrent memory.
We evaluate contiguous sliding-window, approximately uniform, and random selections of the retained layers.

\paragraph{Progressive horizon allocation.}
Recurrent horizons vary progressively with Transformer depth.
Ascending and descending variants allocate increasingly longer horizons toward deeper or shallower layers, respectively.

\paragraph{Two-scale temporal allocation.}
One layer retains the maximum recurrent horizon while the remaining layers retain a shorter recurrent window.
The primary configuration places the full-horizon pathway at the shallowest layer, while the reversed configuration places it at the deepest layer.

\paragraph{Perceiver-AR-like allocation.}
As an additional temporal-pruning reference, we evaluate a Perceiver-AR-inspired configuration~\cite{hawthorne2022perceiverar} in which the shallowest layer retains the maximum recurrent horizon while all higher layers operate only on the current segment.

Together, these baselines test whether preserving complete temporal history only at selected depths is sufficient to maintain functional long-range dependence.

\subsection{KV-Reset Context Utilization Protocol}
\label{sec:exp_krcu}

We evaluate functional dependence on recurrent history using
KV-Reset Context Utilization (KRCU), introduced in
Sec.~\ref{sec:method_krcu}.
For the cross-model KRCU comparison, recurrent KV states are reset at the
8k-token boundary while model parameters remain fixed. We evaluate the
resulting prediction penalty over 22 equally spaced post-reset evaluation
windows. The same reset boundary and evaluation windows are used for all
models to ensure direct comparability.

We report the resulting KRCU curve as a function of distance from the reset
boundary and summarize its magnitude and persistence using KRCU-AUC, as
defined in Eq.~\eqref{eq:krcu_auc}.

In addition to KRCU, we perform a complementary context-restriction
analysis on the full-horizon MHA reference. At evaluation time, we limit
the accessible recurrent history to the most recent 16k, 8k, 4k, or 1k
tokens while keeping model parameters fixed. Target tokens are grouped
according to their absolute positions within the composition. This analysis
measures where distant history contributes to prediction, whereas KRCU
measures how long the predictive effect of previously accumulated recurrent
information persists after a controlled reset.

\subsection{Training Details}
\label{sec:exp_train}

We optimize all models using Adam with the standard Transformer
learning-rate schedule, using 10,000 warmup steps and a peak learning
rate of $3.125\times10^{-4}$.
We take one optimizer step per recurrent segment, with cross-segment
cached states detached from the computation graph.

All models are trained in FP16 without activation checkpointing.
All experiments are conducted on a single consumer-grade NVIDIA GeForce
RTX 4080 GPU with 16\,GB of memory.

\subsection{Evaluation Metrics}
\label{sec:exp_eval}

We report validation perplexity (PPL), training throughput, peak GPU memory,
and wall-clock training cost. Long-range context utilization is evaluated
using the KRCU curve and KRCU-AUC defined in Sec.~\ref{sec:method_krcu}.

\begin{table}[t]
\centering
\caption{
Key model and recurrence configurations.
FHCR reduces the number of KV heads while preserving the full recurrent horizon.
}
\label{tab:hyperparams}
\begin{tabular}{l l}
\hline
Item & Value \\
\hline
Transformer layers $L$ & 18 \\
Hidden size $d$ & 1024 \\
Query heads $H$ & 16 \\
Head dimension $d_{\mathrm{head}}$ & 64 \\
FFN size $d_{\mathrm{ff}}$ & 4096 \\
Segment length $s$ & 1024 \\
Maximum composition length & 32768 \\
Maximum recurrent horizon & 31744 \\
MHA KV heads & 16 \\
FHCR KV heads & 2 \\
\hline
\end{tabular}
\end{table}

\section{Results}
\label{sec:results}

We first establish whether full-horizon recurrence makes functional use of
distant musical context. We then compare temporal-memory pruning with FHCR
in terms of predictive quality, KRCU, and computational cost.

\subsection{Full-Horizon Recurrence Utilizes Distant Musical Context}
\label{sec:results_full_context}

\begin{figure}[t]
  \centering
  \includegraphics[width=\linewidth]{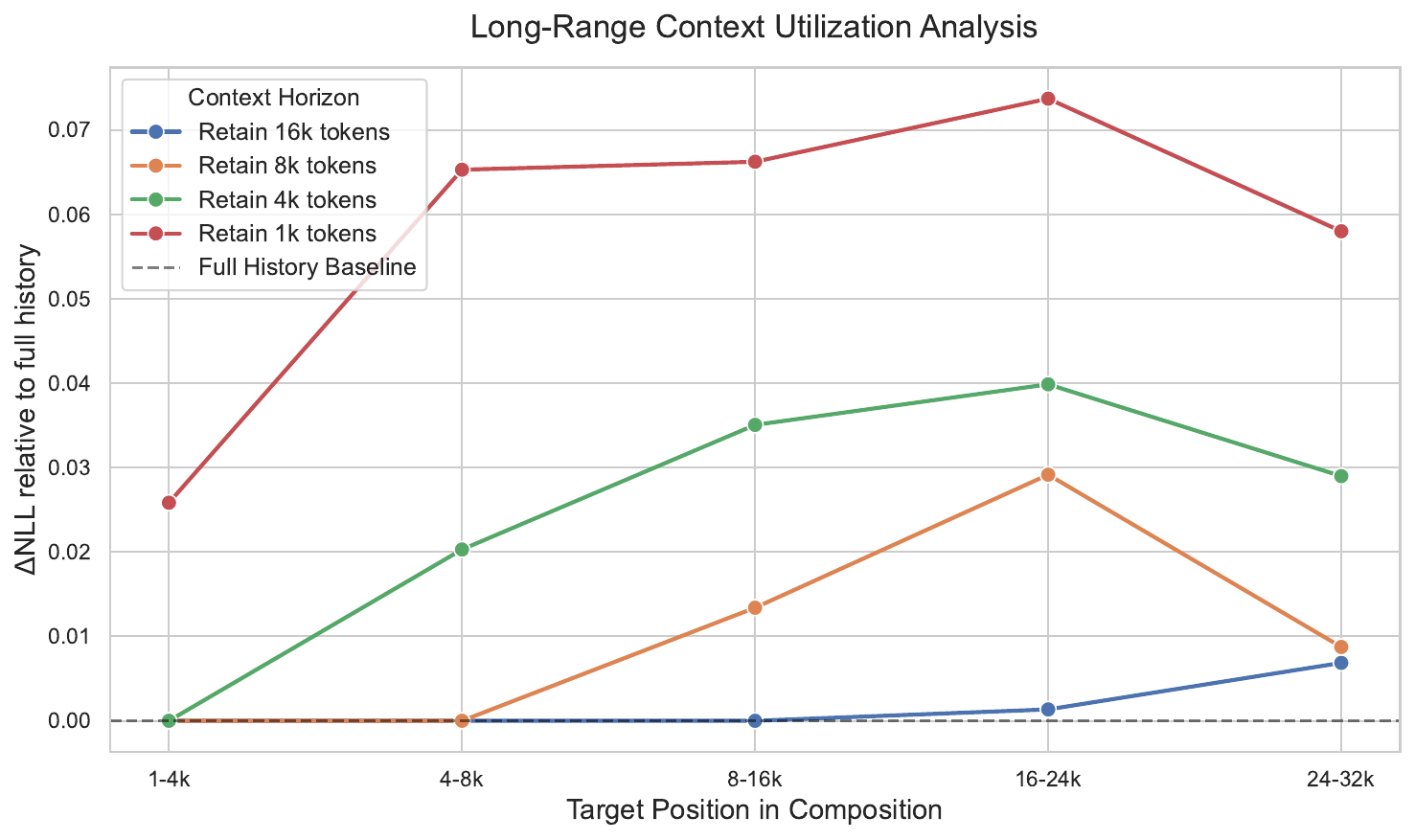}
  \caption{
  \textbf{Long-range context utilization in the full-horizon MHA reference.}
  Increase in negative log-likelihood when the accessible recurrent history is restricted to the most recent 16k, 8k, 4k, or 1k tokens.
  Targets are grouped by their absolute positions within the composition.
  Positive $\Delta$NLL indicates that removed earlier history contributed to prediction.
  }
  \label{fig:full_memory_context_utilization}
\end{figure}

Using the full-horizon MHA recurrent reference, we restrict the accessible
recurrent history at evaluation time while keeping the trained model parameters
fixed. Figure~\ref{fig:full_memory_context_utilization} reports the resulting
increase in negative log-likelihood relative to unrestricted full-horizon
inference.

The results show clear dependence on musical context extending substantially beyond the 1024-token recurrent segment.
For targets between 4k and 8k tokens, restricting history to 4k already produces a measurable NLL increase.
The effect becomes progressively stronger for later target positions and shorter retained histories.
For targets in the 16k--24k region, restricting history to 8k, 4k, and 1k tokens produces increasingly large prediction penalties.

Importantly, even restricting the recurrent horizon to 16k tokens produces measurable degradation for targets near the end of the longest compositions.
Thus, predictions can benefit from musical information originating more than 16k tokens earlier.

These results establish that full-horizon recurrence is not merely storing distant recurrent states.
Information far outside the local segment window contributes measurably to prediction, providing direct empirical motivation for preserving complete-composition context during whole-piece training.

\subsection{Temporal Pruning Preserves Perplexity but Weakens Long-Range Context Utilization}
\label{sec:results_temporal}

We next examine whether reducing recurrent memory along the temporal
dimension preserves both aggregate predictive performance and functional
dependence on distant history.

\begin{figure*}[t]
  \centering
  \includegraphics[width=\textwidth]{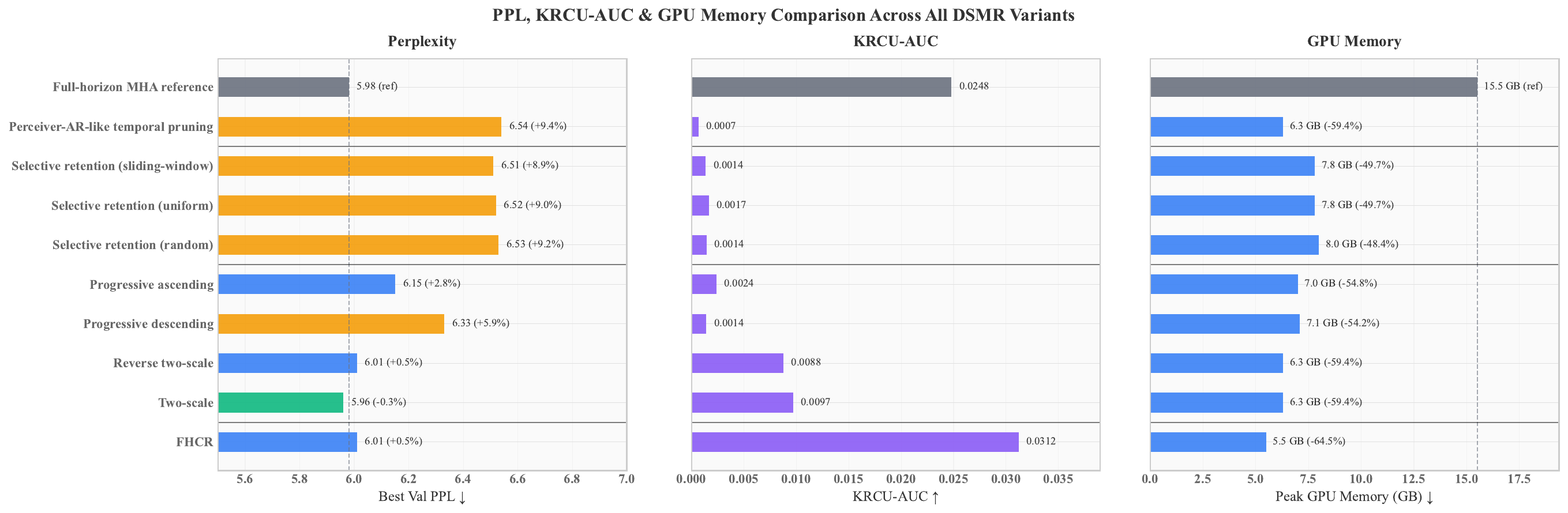}
  \caption{
  \textbf{Comparison of modeling quality, long-range context utilization,
  and memory cost across recurrent-memory configurations.}
  Left: best validation perplexity (lower is better).
  Middle: KRCU-AUC following the 8k KV reset (higher indicates a larger and more persistent predictive effect of pre-reset recurrent information).
  Right: peak GPU memory during training (lower is better).
  FHCR closely matches the full-horizon MHA reference in perplexity,
  exhibits a strong KRCU response, and substantially reduces peak GPU memory,
  whereas temporal-memory-pruning configurations generally show much weaker
  long-range context utilization.
  }
  \label{fig:mem_alloc}
\end{figure*}

Figure~\ref{fig:mem_alloc} provides an overall comparison across three
evaluation dimensions: validation perplexity, KRCU-AUC, and peak training
memory.

Several temporal-memory-pruning configurations remain competitive in
aggregate predictive performance. The two-scale configuration provides
the clearest example, achieving a Best Val PPL of 5.96 compared with
5.98 for the full-horizon MHA reference. Reverse two-scale similarly
reaches 6.01. These configurations also substantially reduce training
memory and increase throughput.

Table~\ref{tab:efficiency} summarizes the corresponding efficiency results.

\begin{table*}[t]
  \centering
  \caption{
  Efficiency metrics for the full-horizon MHA reference,
  temporal-memory-pruning baselines, and FHCR on MAESTRO.
  Percentages in parentheses are relative to the full-horizon MHA reference.
  }
  \label{tab:efficiency}
  \small
  \setlength{\tabcolsep}{6pt}
  \begin{tabular}{l c c c c}
    \hline
    Setting & Tokens/s & Peak GPU Mem (GB) &
    Time to best checkpoint (h) & Best Val PPL \\
    \hline
    Full-horizon MHA reference
      & 7{,}618 & 15.5 & 6.06 & 5.98 \\
    Perceiver-AR-like temporal pruning
      & 11{,}753 (+54.3\%) & 6.3 (-59.1\%) & 3.92 (-35.3\%) & 6.54 (+9.4\%) \\
    \hline
    Selective retention (sliding-window)
      & 11{,}061 (+45.2\%) & 7.8 (-49.6\%) & 4.17 (-31.1\%) & 6.51 (+9.0\%) \\
    Selective retention (uniform)
      & 11{,}257 (+47.8\%) & 7.8 (-49.3\%) & 4.10 (-32.3\%) & 6.52 (+9.1\%) \\
    Selective retention (random)
      & 11{,}160 (+46.5\%) & 8.0 (-48.0\%) & 4.13 (-31.9\%) & 6.53 (+9.3\%) \\
    \hline
    Progressive ascending
      & 10{,}540 (+38.4\%) & 7.0 (-54.5\%) & 4.38 (-27.7\%) & 6.15 (+2.9\%) \\
    Progressive descending
      & 10{,}556 (+38.6\%) & 7.1 (-54.3\%) & 4.37 (-27.9\%) & 6.33 (+5.9\%) \\
    Reverse two-scale
      & 10{,}324 (+35.5\%) & 6.3 (-59.1\%) & 4.47 (-26.2\%) & 6.01 (+0.5\%) \\
    Two-scale
      & 10{,}339 (+35.7\%) & 6.3 (-59.1\%) & 4.46 (-26.4\%) & 5.96 (-0.3\%) \\
    \hline
    FHCR
      & 8{,}982 (+17.9\%) & 5.5 (-64.5\%) & 5.24 (-13.5\%) & 6.01 (+0.5\%) \\
    \hline
  \end{tabular}
\end{table*}

Aggregate perplexity, however, conceals substantial differences in how
these models use recurrent history. We therefore examine their KRCU
responses after resetting recurrent KV states at the 8k-token boundary.

\begin{figure*}[t]
  \centering
  \includegraphics[width=\textwidth]{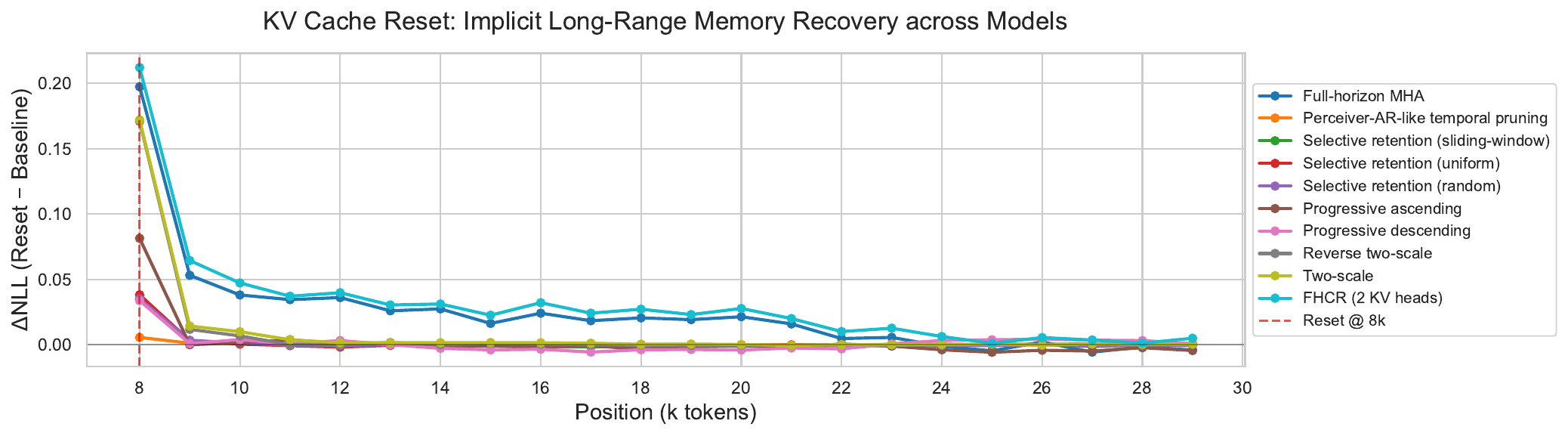}
  \caption{
  \textbf{KV-Reset Context Utilization (KRCU) across recurrent-memory
  configurations.}
  Recurrent KV states are reset at the 8k-token boundary (red dashed line).
  The vertical axis shows the increase in NLL relative to uninterrupted
  recurrent inference.
  Persistent positive $\Delta$NLL indicates that information accumulated
  before the reset boundary continued to contribute to later predictions.
  }
  \label{fig:kv_reset_all}
\end{figure*}

Figure~\ref{fig:kv_reset_all} shows the resulting recovery trajectories.

The models exhibit sharply different recovery behavior.
The full-horizon MHA reference shows a large immediate prediction penalty
after the reset followed by a long recovery tail that remains positive
across many subsequent evaluation windows.
This indicates that information accumulated before 8k continues to affect
predictions far downstream.

Most temporal-pruning models instead recover rapidly toward their
uninterrupted baselines.
Although many exhibit a short-lived increase in NLL immediately after the
reset, their $\Delta$NLL values approach zero after relatively few
post-reset windows.
Their predictions therefore become largely insensitive to information
that existed before the reset boundary once recent context has been
re-accumulated.

The difference is particularly striking for the two-scale model.
Despite achieving a slightly lower validation perplexity than the
full-horizon reference (5.96 vs.\ 5.98), its KRCU response is substantially
weaker and shorter lived.

We quantify these differences using KRCU-AUC over the same 22 post-reset
evaluation windows, as reported in Table~\ref{tab:krcu_auc}.

\begin{table}[t]
\centering
\caption{
KRCU-AUC after recurrent KV reset at the 8k-token boundary.
Higher values indicate a larger and more persistent prediction penalty
following removal of pre-reset recurrent information.
All models are evaluated over the same 22 post-reset windows.
}
\label{tab:krcu_auc}
\small
\begin{tabular}{l c}
\hline
Model & KRCU-AUC \\
\hline
Full-horizon MHA & 0.024804 \\
Perceiver-AR-like temporal pruning & 0.000657 \\
\hline
Selective retention (sliding-window) & 0.001356 \\
Selective retention (uniform) & 0.001652 \\
Selective retention (random) & 0.001436 \\
\hline
Progressive ascending & 0.002376 \\
Progressive descending & 0.001408 \\
Reverse two-scale & 0.008771 \\
Two-scale & 0.009695 \\
\hline
FHCR (2 KV heads) & \textbf{0.031197} \\
\hline
\end{tabular}
\end{table}

The full-horizon MHA reference obtains a KRCU-AUC of 0.0248, whereas
two-scale and reverse two-scale decrease to 0.0097 and 0.0088,
respectively. The remaining temporal-pruning configurations exhibit still
smaller values, ranging from 0.0004 to 0.0024.

Thus, temporal pruning can preserve competitive aggregate perplexity while
substantially weakening functional dependence on distant recurrent history.
Nominal access to long history at selected layers is therefore insufficient
to reproduce the utilization behavior of full-horizon recurrence.

\subsection{FHCR Preserves Full-Horizon Context Utilization Efficiently}
\label{sec:results_fhcr}

We next examine whether FHCR preserves this long-range utilization while
reducing recurrent-state cost. Following the reset at 8k, FHCR exhibits a large immediate NLL increase
and a persistent recovery tail extending across the remainder of the
composition. Its KRCU-AUC is 0.0312, compared with 0.0248 for the
full-horizon MHA reference.

We do not interpret the larger KRCU-AUC as evidence that FHCR has
intrinsically greater long-range memory capacity than MHA.
KRCU measures functional sensitivity to the removal of accumulated
recurrent information, and a larger response may reflect stronger
dependence rather than greater absolute representational capacity.
The important result is that reducing the number of KV heads does not
eliminate the persistent long-range dependence observed under
full-horizon recurrence.

In contrast to the temporal-pruning models, FHCR retains a persistent KRCU
response despite its substantially lower recurrent-state cost. This indicates
that temporal-coverage reduction and KV-representation reduction have markedly
different effects on functional long-range context utilization.

Compared with the full-horizon MHA reference, FHCR reduces peak GPU memory
from 15.5\,GB to 5.5\,GB (64.5\%) while increasing training throughput from
7,618 to 8,982 tokens/s (17.9\%). Its validation perplexity remains close to
the reference (6.01 vs.\ 5.98), while exhibiting a comparably strong KRCU
response (KRCU-AUC 0.0312 vs.\ 0.0248). These results show that FHCR reduces
the cost of full-horizon recurrence without eliminating measurable long-range
context utilization.

\section{Discussion}
\label{sec:discussion}

\subsection{Why Whole-Piece Training Matters}

The central motivation for whole-piece training is that symbolic music contains
dependencies whose natural scope can extend far beyond a local segment.
Our experiments confirm that these dependencies are not merely architectural
possibilities: models trained with full-horizon recurrence make functionally
measurable use of distant musical history.

This distinction matters because excerpt-based training changes the statistical
training unit itself. Once a composition is divided into independently optimized
windows, dependencies crossing those boundaries are unavailable during learning.
Whole-piece training instead preserves the composition as a continuous context,
while using recurrent segmentation only as a computational device.

The implication is that long-context symbolic music modeling should consider
not only how large a context an architecture can nominally support, but also
whether the training procedure preserves the natural boundaries of the musical
work.

\subsection{Perplexity Is Not a Sufficient Long-Context Metric}

A central implication of our experiments is that aggregate predictive quality
and functional long-range dependence are not equivalent.
Perplexity averages prediction loss over all tokens, while genuinely distant
musical relationships may affect only a comparatively small subset of events.
A model can therefore rely more heavily on local context and still preserve
similar aggregate likelihood.

This creates a potential evaluation blind spot for long-context models:
architectural access to a long history and competitive perplexity do not
establish that distant information is actually influencing predictions.
KRCU addresses this gap by intervening directly on accumulated recurrent state
and observing the persistence of its predictive effect.

We therefore view KRCU as complementary to perplexity rather than a replacement
for it. Perplexity measures overall modeling quality, whereas KRCU probes
functional dependence on information carried across long temporal distances.

\subsection{Preserving Temporal Coverage versus Compressing Representation}

The contrast between temporal pruning and FHCR suggests that recurrent-memory
efficiency has at least two distinct dimensions: the temporal extent of retained
history and the cost of representing each retained state.
Reducing either dimension lowers memory usage, but the two operations need not
have the same effect on learned long-range behavior.

Our results suggest that temporal coverage is particularly important for
whole-piece symbolic music modeling.
Shortening or selectively removing recurrent history can alter the effective
temporal behavior of the model, whereas reducing the KV representation while
preserving the full horizon can retain functional dependence on distant context.

This motivates a broader design principle:
when long-range dependence is a central modeling objective, it may be
preferable to compress the representation of history before compressing its
temporal extent.
FHCR is one realization of this principle using grouped-query attention, but
the principle itself is more general than the specific mechanism used here.

\subsection{Implications for Long-Context Sequence Modeling}

The distinction between temporal coverage and representation cost may extend
beyond symbolic music.
Many efficient long-context methods reduce cost by shortening windows,
sparsifying temporal access, or selectively retaining historical states.
Our results suggest that such methods should be evaluated not only by aggregate
task performance, but also by whether distant information remains functionally
relevant after compression.

KRCU provides one possible diagnostic for stateful sequence models because it
tests the predictive consequence of removing accumulated history without
changing model parameters.
Applying similar interventions to language, code, or other long-form sequential
domains could help distinguish nominal context capacity from context that is
actually used.

\subsection{Limitations}
\label{sec:limitations}

Our experiments are currently limited to the MAESTRO piano dataset.
We deliberately focus on performance-derived symbolic music because it
preserves expressive timing and dynamics and provides a natural pathway
toward substantially larger training corpora. In particular, advances in
automatic music transcription (AMT) may enable performance-level symbolic
representations to be recovered from the vast collections of existing
music recordings, potentially allowing whole-piece models to be trained
on musical material far beyond currently available symbolic datasets.
At present, however, large-scale, high-quality performance-level symbolic
datasets suitable for this setting remain limited, making MAESTRO a
practical choice for the present study.

A further limitation concerns the interpretation of long-range context
utilization. The strong KRCU response of FHCR provides evidence that the
model functionally uses distant context, supporting its architectural
capacity for long-range dependency modeling. However, the quality of
long-range structure in free generation may be further shaped by subsequent
post-training. Accordingly, our evaluation focuses on context utilization
in the pretrained models rather than the quality of their free generations,
although we provide generated samples for qualitative inspection.
Moreover, evaluating long-range structure in generated music remains
challenging, with no universally accepted objective or subjective metric.
KRCU should therefore be interpreted as a diagnostic of functional context
utilization rather than a complete measure of musical generation quality.

\section{Conclusion}
\label{sec:conclusion}

This paper formulated \textbf{whole-piece training} for symbolic music
language models, preserving each complete composition as a continuous
training instance through streamed recurrence. Our experiments show that
full-horizon models make measurable use of musical information far beyond
the local segment window.

We introduced \textbf{KV-Reset Context Utilization (KRCU)} to diagnose
functional long-range dependence and showed that competitive perplexity can
coexist with substantially weakened use of distant recurrent history.
Motivated by this finding, we proposed \textbf{Full-Horizon Compressed
Recurrence (FHCR)}, which reduces recurrent-state cost through fewer KV heads
while preserving the complete temporal horizon and a strong long-range KRCU
response.

These results distinguish two dimensions of recurrent-memory efficiency:
how far history is retained and how that history is represented.
For whole-piece symbolic music modeling, preserving temporal coverage while
reducing representation cost provides a practical path toward memory-efficient
long-context training.

\footnotesize

\section*{Acknowledgements}
\noindent\textbf{AI-Assisted Language Editing.}
During the preparation of this work the authors used ChatGPT in order to improve English grammar and readability. After using this tool, the authors reviewed and edited the content as needed and take full responsibility for the content of the published article.

\bibliographystyle{elsarticle-num}
\bibliography{ref}

\end{document}